\documentclass[%
 reprint,
 amsmath,amssymb,
 aps,
]{revtex4-2}

\usepackage{graphicx}
\usepackage{dcolumn}
\usepackage{bm}
\usepackage{soul,color}

\begin{document}

\preprint{APS/123-QED}

\title{A Microwave Anapole Source Based on Electric Dipole Interactions Over a Low-Index Dielectric}

\author{Muhammad Rizwan Akram}
 \email{muhammadrizwan.akram@utoledo.edu}
\author{Abbas Semnani}%
 \email{abbas.semnani@utoledo.edu}
\affiliation{%
 Department of Electrical Engineering and Computer Science \\
 The University of Toledo
}%

\date{\today}

\begin{abstract}
The pursuit of non-radiating sources and radiation-less motion for accelerated charged particles has captivated physicists for generations. Non-radiating sources represent intricate current charge configurations that do not emit radiation beyond their source domain. In this study, we investigate a single non-radiating source, comprising a low-index dielectric disk excited by a split ring resonator. Employing analytical and numerical methods, we demonstrate that this configuration supports an anapole state, exhibiting minimal or no radiation, effectively representing a non-radiating source. The radiation suppression is accomplished through the destructive interference of electric dipoles excited on the metallic and dielectric components of the proposed prototype. Transforming the design into a cost-effective device capable of suppressing radiation, we achieve impressive numerical and experimental agreement, affirming the formation of the anapole state using the lowest order multi-poles. Moreover, the devised anapole device is remarkably compact, constructed from a low-index dielectric, and employs readily available components. As a versatile platform, the proposed device can spearhead anapole research for diverse applications, including sensing, wireless charging, RFID tags, and other non-linear applications.    
\end{abstract}

\maketitle

It is well known that the accelerating charges emit electromagnetic (EM) radiation in far-field to preserve the stability of matter, composed of atoms and molecules.  This insight played a crucial role in Bohr's formulation of his renowned postulates, eventually laid the groundwork for quantum mechanics. However, from early days, scientists have been trying to find the confined charge-current configurations that don't radiate. The oscillatory motion of charged sphere within a single period is one such configuration suggested by Bohm \cite{a1bohm}, which was later generalized for electron motion in orbits \cite{a2goedecke}. In this context, a particular charge configuration known as the "Anapole" was introduced in elementary particle physics by Yakov Zel'dovich \cite{a3zel}. However, the experimental detection of the anapole effect remained challenging until Wood et al. successfully observed and measured it in Cesium atom through parity-violating effects \cite{a4wood}.

The electrodynamics analogue of an anapole or non-radiating (NR) is achieved through the co-location of fundamental electric and magnetic dipoles, along with their toroidal counterparts. This spatial arrangement leads to far-field destructive interference, resulting in minimal or negligible radiation due to their similar, but out-of-phase, field distributions \cite{b1yang}. The toroidal dipole emerges as the third-order term in the Taylor expansion of electromagnetic potentials, complementing other essential dipole moments defined in both Cartesian and spherical harmonics representations \cite{a5gurvitz,a6alaee}. The toroidal dipole was experimentally realized in 2010, utilizing the unique response of a metamaterial \cite{a7kael}. In 2013, an anapole was observed experimentally using metasurfaces under plane wave excitation in the microwave spectrum \cite{a8fedotov}. Subsequently, in 2015, a simple silicon disk was employed to demonstrate an optical anapole \cite{a9miro}.

Due to their exceptional characteristics, including near-field enhancement, high-quality factor (Q), and far-field suppression, anapoles have garnered significant attention. This heightened interest has led to a series of advancements in anapole technology \cite{a5gurvitz,a11Kapitanova, a12luk, a13luk, a14parker, a15zenin}, along with the introduction of new applications in sensing \cite{a16zhang}, power transfer \cite{a17zanganeh}, and quantum technologies \cite{a18savinov}. However, these developments are primarily limited to plane wave excitations. Hence, expanding the scope of anapole technology to accommodate other excitation methods could unlock additional opportunities for innovative applications beyond the current plane wave structures.

Recently, a novel single anapole source was successfully demonstrated, employing a dipole surrounded by four high refractive index cylinders \cite{a19nemkov}. In this configuration, the central dipole excites an electric toroidal dipole in the surrounding rods, leading to destructive interference and the formation of the anapole state. In another study by Esmaeel et al. \cite{a20zanganeh}, anapole formation was achieved using a high refractive index cylinder, excited either by an electric dipole with a metallic rod in the middle or by a magnetic dipole with a loop placed inside the cylinder. In the former approach, the superposition of electric dipoles and electric toroidal dipoles forms the anapole state, while in the latter approach, only magnetic dipoles are superimposed to create the anapole configuration. In a different study, an anapole state was achieved using a high refractive index disk excited by an external loop, which led to the superposition of electric dipoles and quadrupoles \cite{a21zanganeh}. However, all these works relied on high refractive index materials, which were custom-made and not commercially available.

As an alternate approach, recent works have revealed that it is possible to achieve an anapole state without involving toroidal dipoles. For instance, anapole states have been demonstrated using only electric dipoles and quadrupoles \cite{a21zanganeh}. Numerical studies \cite{a20zanganeh} have explored the superposition of only magnetic dipoles, while theoretical investigations \cite{a22zurita} have depicted the possibility of using superimposed electric dipoles alone to achieve the anapole state. However, to date, no experimental demonstrations have been conducted to verify the realization of anapole states solely through the use of basic electric or magnetic dipoles.

In this letter, we present a novel anapole source composed of a commercially-available and low-index dielectric cylinder, coupled with a simple excitation topology. This study experimentally proves that if combined with a loop as a radiation source, a low-index dielectric cylinder can effectively showcase the anapole state. By analyzing the Cartesian multi-pole expansion in the long wavelength approximation \cite{a6alaee}, we confirm the feasibility of anapole formation at the lowest order using only dipole-dipole interactions, as previously explored in theoretical studies \cite{a22zurita}. For experimental realization, a dielectric cylinder is placed on a metallic plate, excited by a microstrip line coupled through an etched slot on the metallic plate. This configuration results in a significant size reduction compared to existing anapole sources \cite{a20zanganeh,a21zanganeh,a19nemkov}, makes it compatible with the printed circuit board (PCB) fabrication technology, and enables effective impedance matching, a challenge often encountered in anapole-based designs.

The novelty of the proposed design lies in its reliance solely on the lowest-order electric dipoles, enabling a compact size even with a low dielectric constant of, for example, $\epsilon_r$ = 13, in contrast to earlier works that used $\epsilon_r$ in the range of 1000 \cite{a19nemkov,a20zanganeh,a21zanganeh}. Moreover, its compatibility with existing fabrication technologies allows for commercial prototyping and facilitates the rapid development of various potential anapole-based applications in wireless sensing, charging, and non-linear electromagnetics/optics.

\begin{figure}[t]
\includegraphics[width=\linewidth]{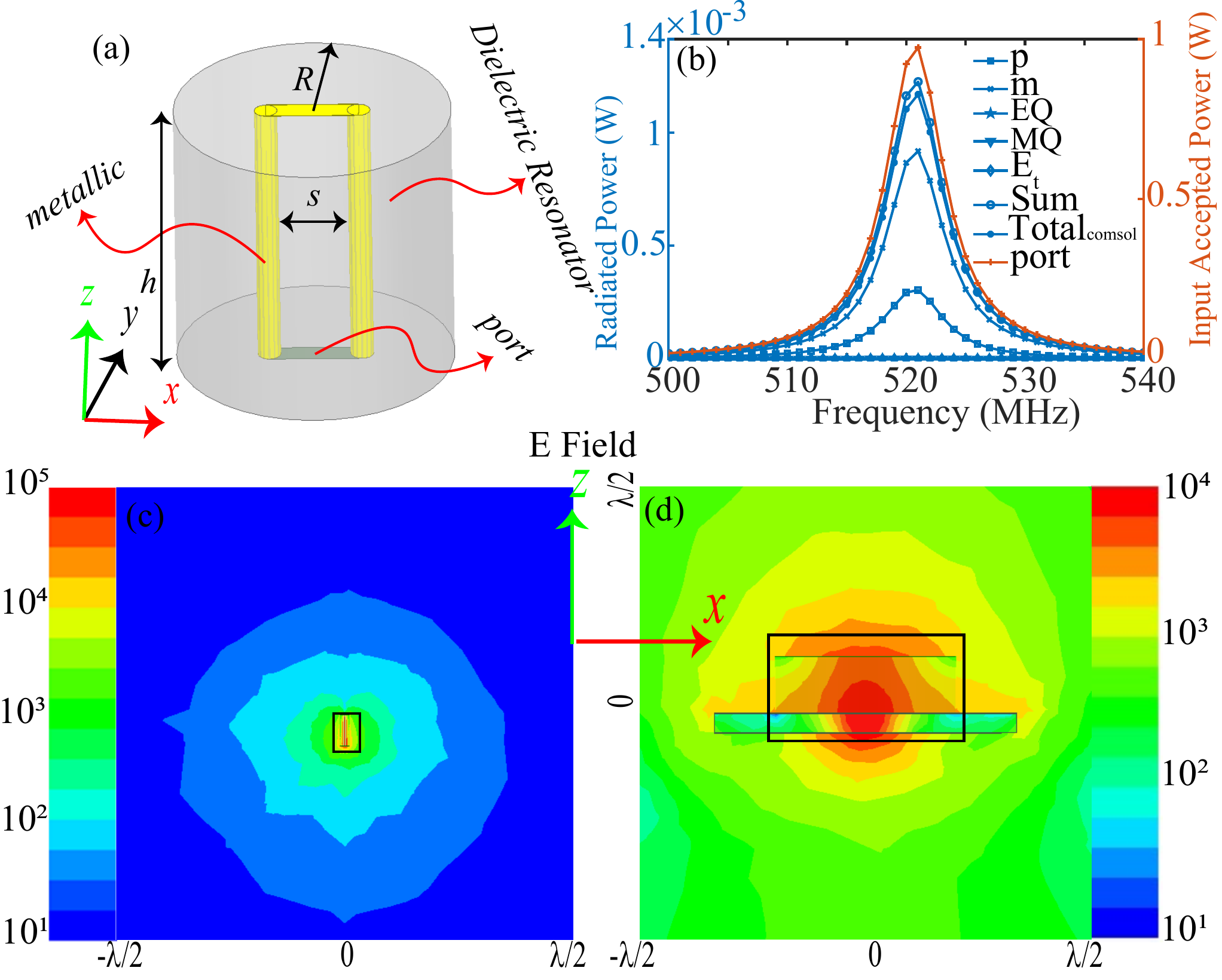}
\caption{\label{fig:fig1}(a) Schematic of the proposed anapole design with a dielectric cylinder of radius \textit{R} = 6 mm and height \textit{h} = 40 mm, with metallic vias each of radius r = 0.5 mm, with center to center distance of \textit{s} = 0.6 mm, such that the top ends are connected, forming a split ring resonator (SRR). The dielectric cylinder is made of Rogers TMM13 laminate with permittivity of 12.85 and tan~$\delta = 1.9\times 10^{-3}$. (b) The radiated power evaluated through numerical Cartesian multi-pole expansion, and the total power accepted by the device in red. The simulated near electric field around the design (c) the proposed non-radiating (anapole) source, and (d) $HE11\delta$ cylindrical dielectric resonator radiating source.}
\end{figure}

The proposed anapole design, depicted in Fig.~\ref{fig:fig1}, consists of two metallic rods connected by a metallic strip, configured to form a loop. This loop radiates like an electric dipole and induces a nearly equal but opposite electric dipole in the surrounding dielectric cylinder. As a result, the two dipoles destructively interfere, giving rise to an anapole state. The design's non-radiating response is highly sensitive to the gap between the two metallic rods. Consistent with theoretical investigations \cite{a22zurita}, a narrower gap leads to lower radiation, making it an essential parameter to tailor the design for specific applications.  

The radiation performance of the proposed anapole design is initially explored through numerical simulations using COMSOL multi-physics, as depicted in Fig.\ref{fig:fig1}(b). At the resonant frequency of 520 MHz of a sample design, the maximum radiation amounts to just 1.4 mW out of the total input power of 1 W, representing only 0.14\% of radiated power. Remarkably, only 0.04\%  of this radiation is attributed to the electric dipoles, and the rest is from the magnetic dipole component, as plotted in detail in Fig.\ref{fig:fig1}(b).

For radiation comparison, the HE11$\delta$ radiating mode of a cylindrical dielectric resonator antenna (DRA) operating at 5.05 GHz is used as a reference, due to similarity in field distribution. The near-field plots of the XZ plane for both non-radiating and radiating cases are shown in Figs.~\ref{fig:fig1}(c) and (d), respectively. From Fig.~\ref{fig:fig1}(c), it is evident that the electric field is highly concentrated between the two metallic rods, and it is effectively suppressed around the entire device in contrast to the radiating state, shown in Fig.~\ref{fig:fig1}(d). Furthermore, the near-field intensity for the non-radiating mode is observed to be more than 100 times higher compared to the radiating state. The remarkable confinement of electric and magnetic fields in the proximity of the device, coupled with the significant suppression of far-field radiation provides compelling evidence of the successful formation of an anapole state.

\begin{figure}[h]
\includegraphics[width=\linewidth]{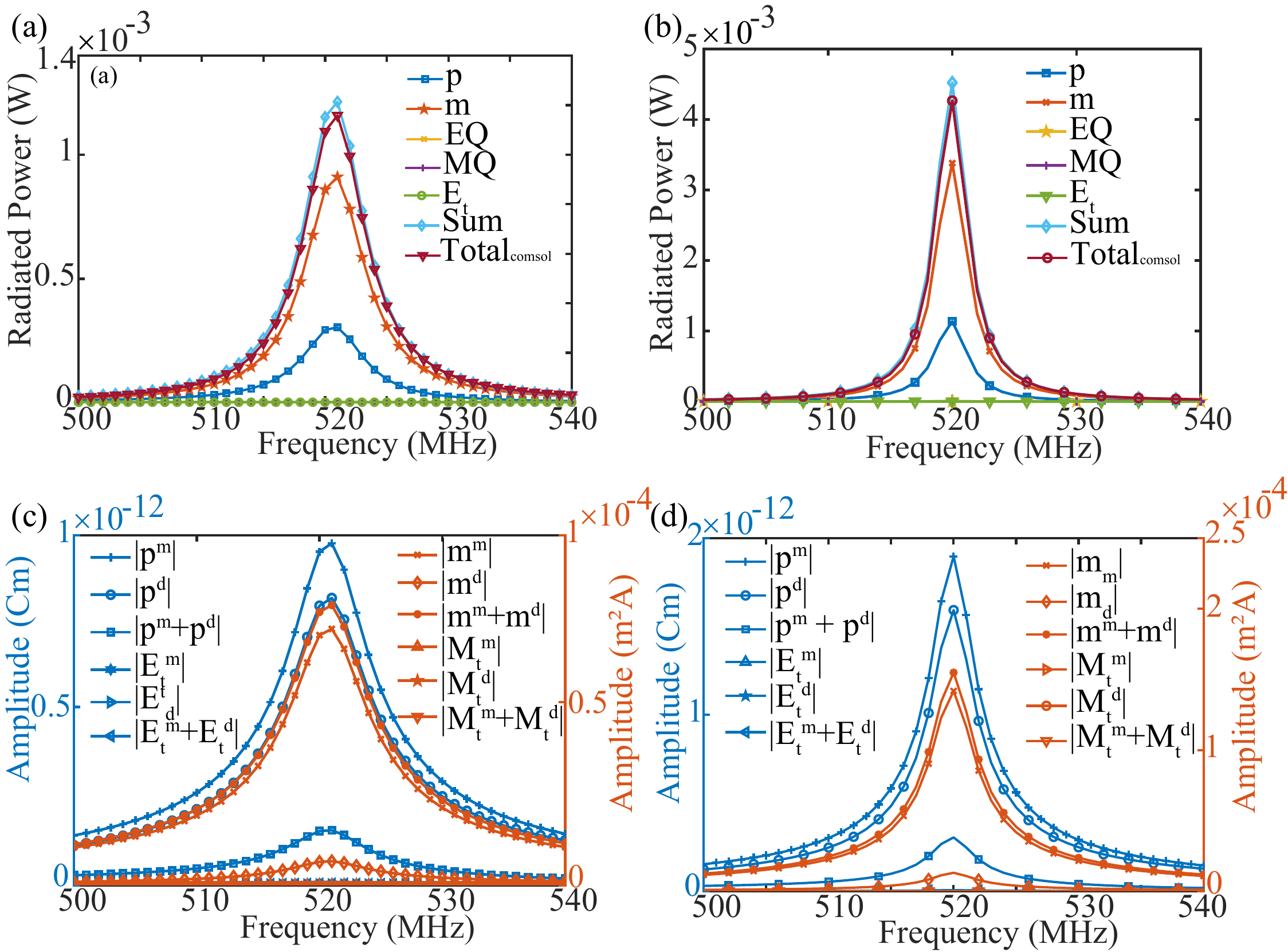}
\caption{\label{fig:fig2}Extracted radiated powers using Cartesian multi-pole expansion and numerically for vias separation of (a) \textit{s} = 0.6 mm and (b) 1.2 mm, while input impedance ranges from 15k to 150k$\Omega$, depending on \textit{s}. Similarly, electric and magnetic dipoles evaluation on loop and dielectric disk separately for (c) \textit{s} = 0.6 mm and (d) \textit{s} = 1.2 mm.}
\end{figure}

To gain a profound understanding of the non-radiating response's underlying mechanism, a Cartesian multi-pole expansion of the EM fields is conducted. The results are depicted in Fig.~\ref{fig:fig2} for two different separations between the metallic rods, \textit{s}. Notably, Figs.~\ref{fig:fig2}(a,b) clearly demonstrate that a smaller \textit{s} leads to more pronounced radiation suppression. This observation aligns with the theoretical predictions for analogous cases of electric dipole interactions \cite{a22zurita}. As radiation contributions are solely attributed to electric and magnetic dipoles, separate characterizations of these dipoles on both the metallic loop and the dielectric section are performed, with the results illustrated in Figs.~\ref{fig:fig2}(c,d). These results reveal that the electric dipoles induce a destructive interference on the loop, similar to that observed in the dielectric region, whereas the magnetic fields exhibit constructive interference. Additionally, the toroidal dipoles contribute almost negligibly to the overall response. Consequently, the expression for this non-radiating electric source can be represented as \cite{a23evlyukhin}:
\begin{equation}
    P  \sim  | p^m + p^d |^2 = (p^m)^2 + (p^d)^2 + 2RE[p^mp^d*],
\end{equation}
where $p^m$ is the electric response of the metallic loop, $p^d$ represents the electric response of the dielectric cylinder, and $( p^m + p^d )$ is the total electric response of the structure.

\begin{figure}[h]
\includegraphics[width=\linewidth]{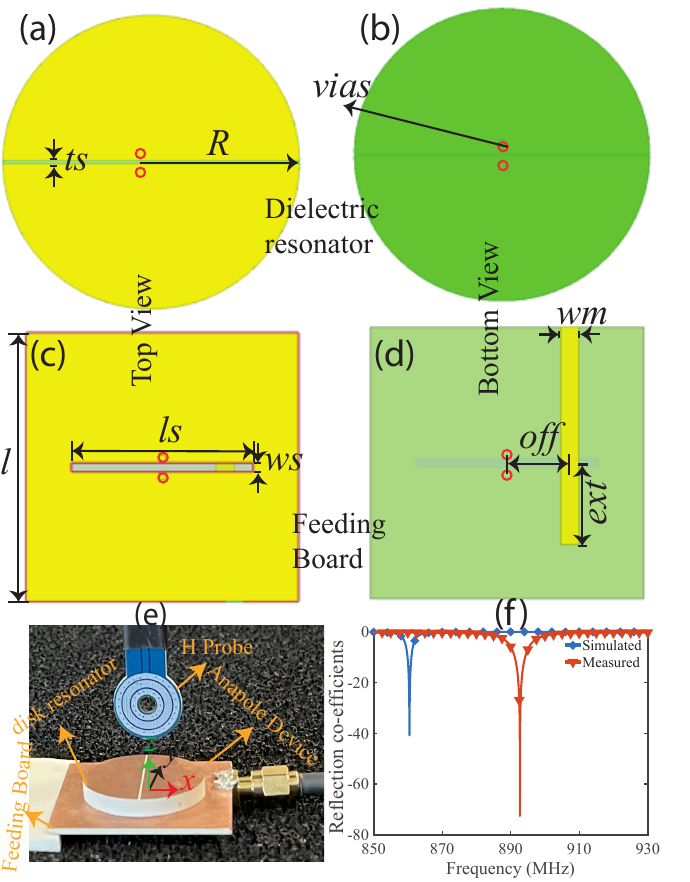}
\caption{\label{fig:Fig3}Top (a,c) and bottom (b,d) views of the dielectric resonator and the feeding board of the anapole device with \textit{R} = 14 mm, thickness of the dielectric cylinder \textit{h} = 3.81 mm, thickness of feed board $h_2$ = 1.27 mm, $l_s$ = 20 mm, $l$ = 40 mm, $w_m$ = 2 mm, $ext$ = 9 mm, $off$ = 6.9 mm, $w_s$ = 1 mm, $t_s$ = 0.3 mm, and $s$ = 1.2 mm. The cylindrical resonator is made of TMM13i with permittivity of $\epsilon_r$ = 13 and $tan\delta$ = $1.9\times10^{-3}$. The bottom board is made of TMM6 with permittivity of $\epsilon_r$ = 6 and $tan\delta$ = $2.3\times10^{-3}$. (e) The fabricated anapole device connected to the transmitter port of a vector network analyzer (VNA) through a 50-$\Omega$ cable, with an H-field probe on top attached to the receiver port of the VNA, configured for $H_y$ measurements. The bottom board is the feeding part which is coupled to the disk using a slot coupling topology. (f) The simulated and measured reflection coefficients, showcasing a solid resonance.}
\end{figure}

As an experimental prototype, a cylindrical disk is meticulously crafted from a commercially available Rogers TMM13i laminate, featuring a thickness of 3.81 mm and 35 $\mu$m copper cladding. For the feeding section, another layer is fashioned from a 1.27-mm thick Rogers TMM6 laminate, with the same 35 $\mu$m cladding. The assembly process involves aligning the disk with the feeding board using vias and holes, which is then firmly secured using silver epoxy. The detailed design is shown in Figs.~\ref{fig:Fig3}(a-e). The EM energy is fed to 50-$\Omega$ microstrip line that is coupled to dielectric resonator through the slot, etched on the ground plane. Fine-tuning the slot's position relative to the microstrip line allows for impedance matching. A top metallic pattern is utilized atop the dielectric resonator for further size reduction. The resulting prototype is remarkably compact, with its largest dimension lower than 0.1$\lambda$. To facilitate field measurements, a monopole and a loop receiver acting as an E and H probes, respectively, are utilized. The detailed design along with fabricated device and the automated near-field measurement system are depicted in Fig.~\ref{fig:Fig6}(a).

\begin{figure}[h]
\includegraphics[width=\linewidth]{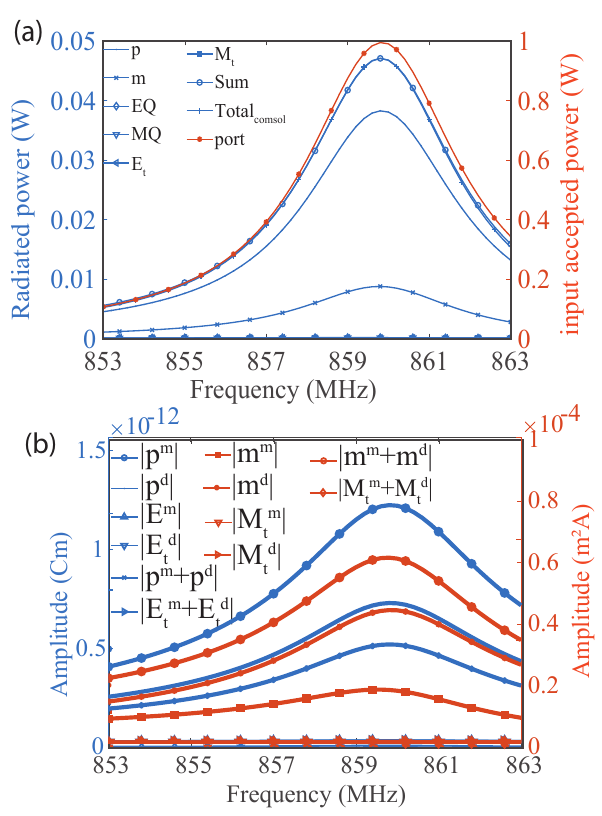}
\caption{\label{fig:Fig4} (a) Cartesian multi-pole expansion and numerical evaluation for radiated and accepted powers, and (b) contribution assessment of the metallic and dielectric parts for electric and magnetic dipoles.}
\end{figure}

The simulated and measured return losses are plotted in Fig.~\ref{fig:Fig3}(f). A strong resonance occurs at approximately 893 MHz, with around a 30-MHz deviation compared to the simulation, attributed to the manual alignment of the disk. Subsequently, a Cartesian multi-pole expansion analysis was conducted showing that less than 50 mW out of 1 W accepted power is radiated, with a contribution of only 35 mW from the electric dipoles as can be seen from Figs.~\ref{fig:Fig4}(a ,b). Further reduction of this radiation is feasible by narrowing the gap between the vias; however, practical limitations pertaining to fabrication tolerances should be considered. Moreover, the individual contributions of electric and magnetic dipoles were evaluated separately on the metallic and dielectric portions of the structure. The experimental results validate the achievement of significant radiation suppression.

\begin{figure}[!h]
\includegraphics[width=\linewidth]{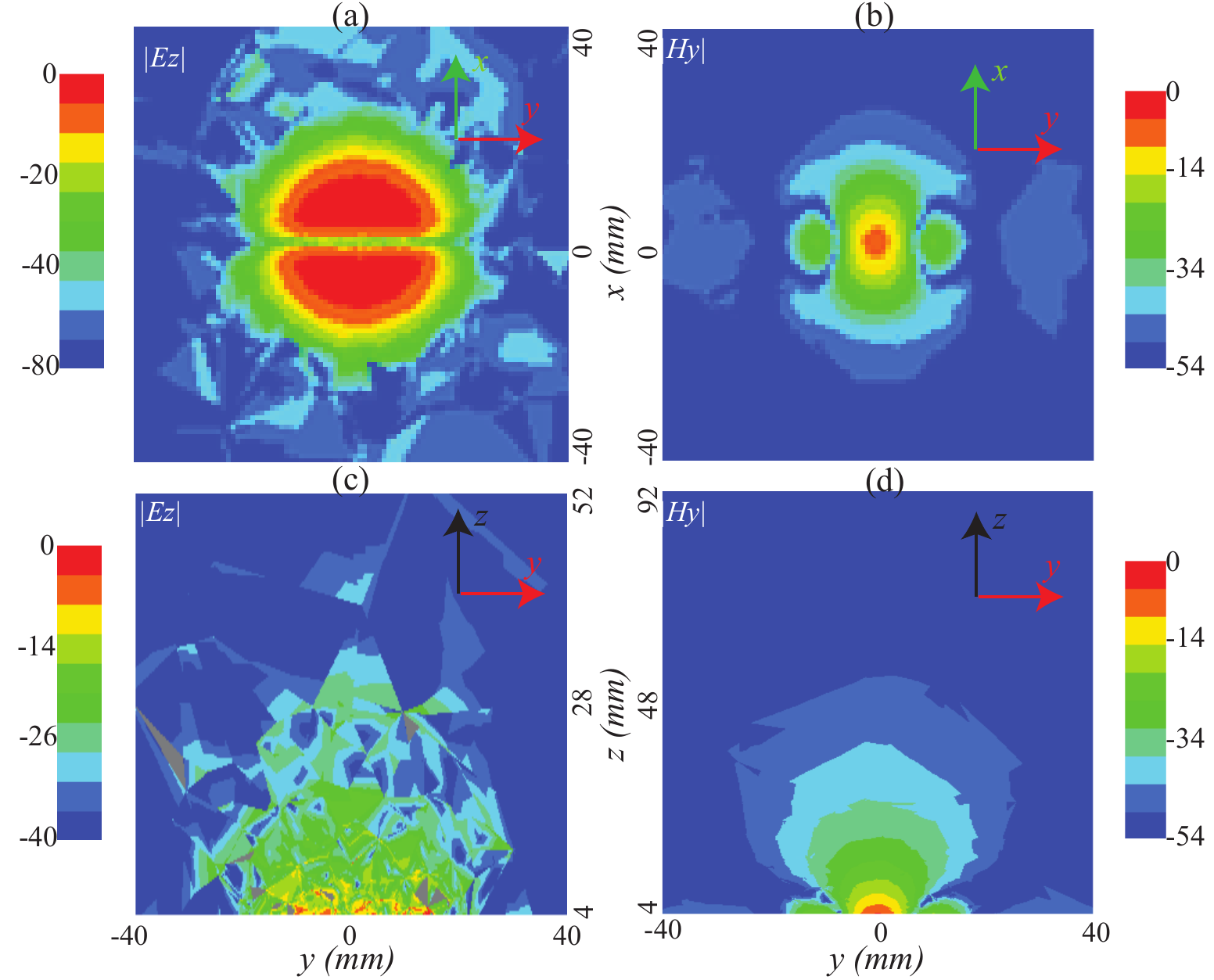}
\caption{\label{fig:Fig5}The simulated $E_z$ for (a) \textit{XY} plane at a height of \textit{z} = 4 mm, (c) \textit{YZ} plane with \textit{x} = 0, and $H_y$ at (b) \textit{XY} plane at a height of \textit{z} = 4 mm, and (d) \textit{YZ} plane with \textit{x} = 0.}
\end{figure}

\begin{figure}[h]
\includegraphics[width=\linewidth]{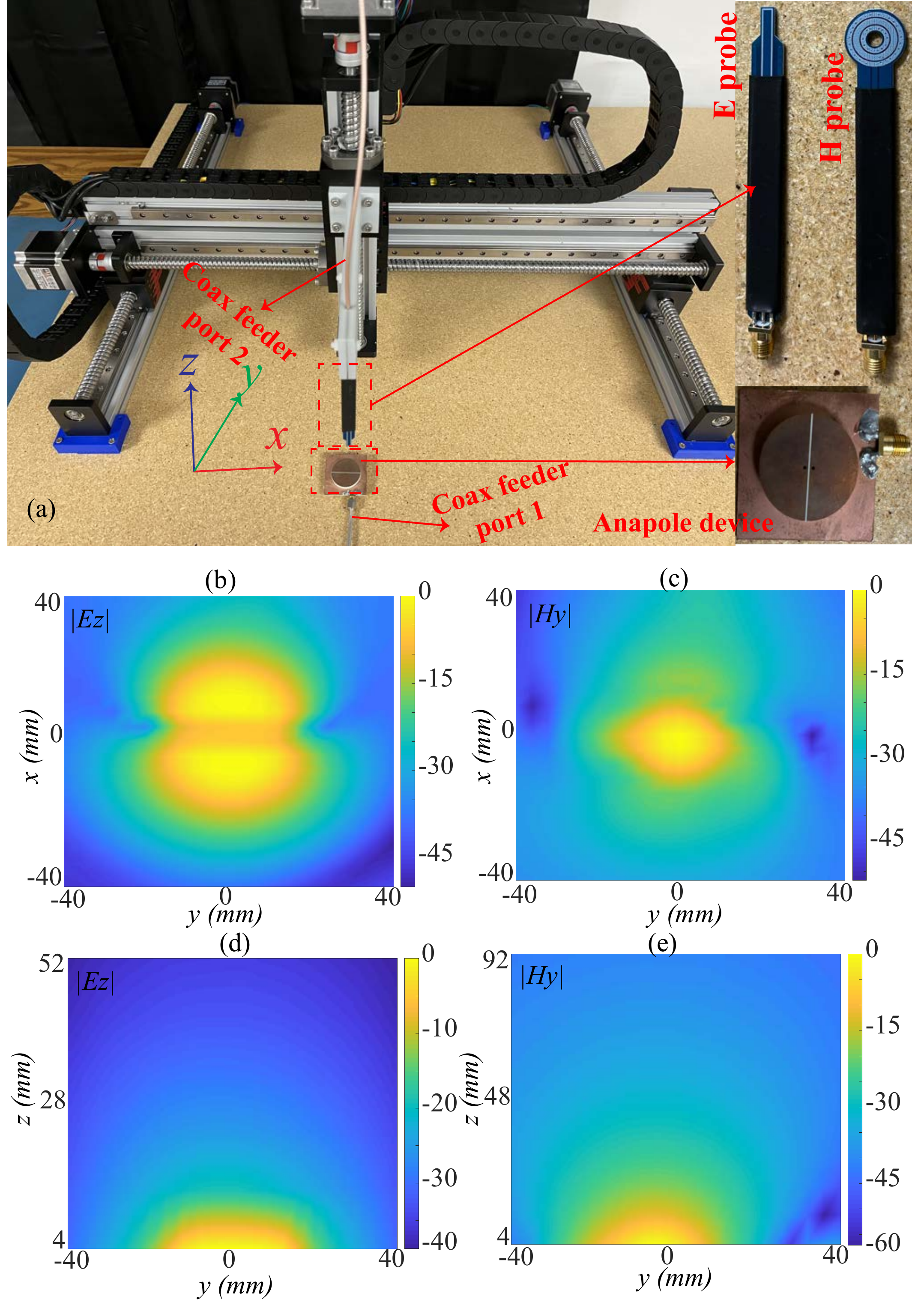}
\caption{\label{fig:Fig6}(a) Automated measurement system along with used probes \& the device and The measured $E_z$ for (b) \textit{XY} plane at a height of \textit{h} = 4 mm, (d) \textit{YZ} plane with \textit{x} = 0, and $H_y$ for (c) \textit{XY} plane at a height of \textit{h} = 4 mm, and (e) \textit{YZ} plane with \textit{x} = 0.}
\end{figure}

The near-field investigation of the prototype is conducted through both numerical simulations and experimental measurements, as depicted in Fig.\ref{fig:Fig5} and Fig.\ref{fig:Fig6}. Among the field components, only $E_x$, $E_z$, and $H_y$ demonstrate significant contributions, while the remaining field components are nearly negligible and can be safely disregarded. For the sake of conciseness, we focus on measuring and comparing only the $E_z$ and $H_y$ components with the simulation results. Further comprehensive details can be found in the supplementary materials \cite{a36}.

To facilitate the measurements, a monopole probe is employed for $E$-field measurements, while a loop-shaped probe is utilized for $H$-field measurements. As shown in Fig.\ref{fig:Fig3} (c), the probes are oriented strategically, aligning their axes along the y-axis to optimally couple with the maximum $H_y$ field and vertically for the monopole probe to couple maximally with $E_z$. The field plots exhibit an excellent agreement between the simulations and measurements, revealing a robust suppression of the fields, confined within a few fractions of a wavelength away from the device. This striking observation serves as a clear indication of the successful formation of the anapole (non-radiating) state. 

 As a concluding remark, we would like to emphasize that the anapole state does not inherently arise as a natural eigenstate within an open cavity, as highlighted in previous literature \cite{a24mon}. It only manifests in the presence of an incident field, setting it apart from bound states in continuum or traditional eigenstates \cite{a25hsu, a26mon, a27hsu, a28silv, a29mon,a30doel}. However, the anapoles are externally excited and sustained states, distinct from scattering effects. The fact that the anapole state results in the cancellation of far-field interaction of various dipolar responses enables them for the near-field enhancement around the device, which can be utilized for multiple applications as compared to cavities in which fields are confined within metallic enclosures. Furthermore, the cloaked devices depicted in \cite{a31alu, a32alu, a33edwards, a34alu, a35bilotti}, which similarly exhibit no external scattering, stand distinct from anapoles due to the underlying realization mechanism. Unlike anapoles, these devices are crafted from multilayered particles, wherein the cumulative dipolar response on each layer nullifies, resulting in a non-scattering cross-section. 

We have introduced a novel concept for the lowest order anapole source, relying solely on electric dipole interactions. To realize this idea, a commercially-available low-index dielectric was employed in a hybrid structure, comprising a dielectric disk inserted with a metallic loop. Through extensive numerical investigations, it was established that effective radiation suppression can indeed be achieved by leveraging electric dipole-electric dipole interactions. To validate the practical feasibility, we proposed and numerically verified a specific design, confirming its capability to achieve the desired radiation suppression through electric dipole interactions. Subsequently, experimental evaluations demonstrated reasonable agreements with the numerical predictions. The proposed device was then experimentally evaluated, and reasonable agreements were found. Notably, the proposed device boasts compactness and utilizes a low-index dielectric, which holds promise for a wide range of anapole applications. These applications encompass various fields, including sensing, wireless power transfer, and non-linear interactions.

This work was supported by the National Science Foundation under Grant ECCS-2102100.

\nocite{*}

\bibliography{refs}

\providecommand{\noopsort}[1]{}\providecommand{\singleletter}[1]{#1}%
\begin{thebibliography}{38}%
\makeatletter
\providecommand \@ifxundefined [1]{%
 \@ifx{#1\undefined}
}%
\providecommand \@ifnum [1]{%
 \ifnum #1\expandafter \@firstoftwo
 \else \expandafter \@secondoftwo
 \fi
}%
\providecommand \@ifx [1]{%
 \ifx #1\expandafter \@firstoftwo
 \else \expandafter \@secondoftwo
 \fi
}%
\providecommand \natexlab [1]{#1}%
\providecommand \enquote  [1]{``#1''}%
\providecommand \bibnamefont  [1]{#1}%
\providecommand \bibfnamefont [1]{#1}%
\providecommand \citenamefont [1]{#1}%
\providecommand \href@noop [0]{\@secondoftwo}%
\providecommand \href [0]{\begingroup \@sanitize@url \@href}%
\providecommand \@href[1]{\@@startlink{#1}\@@href}%
\providecommand \@@href[1]{\endgroup#1\@@endlink}%
\providecommand \@sanitize@url [0]{\catcode `\\12\catcode `\$12\catcode
  `\&12\catcode `\#12\catcode `\^12\catcode `\_12\catcode `\%12\relax}%
\providecommand \@@startlink[1]{}%
\providecommand \@@endlink[0]{}%
\providecommand \url  [0]{\begingroup\@sanitize@url \@url }%
\providecommand \@url [1]{\endgroup\@href {#1}{\urlprefix }}%
\providecommand \urlprefix  [0]{URL }%
\providecommand \Eprint [0]{\href }%
\providecommand \doibase [0]{https://doi.org/}%
\providecommand \selectlanguage [0]{\@gobble}%
\providecommand \bibinfo  [0]{\@secondoftwo}%
\providecommand \bibfield  [0]{\@secondoftwo}%
\providecommand \translation [1]{[#1]}%
\providecommand \BibitemOpen [0]{}%
\providecommand \bibitemStop [0]{}%
\providecommand \bibitemNoStop [0]{.\EOS\space}%
\providecommand \EOS [0]{\spacefactor3000\relax}%
\providecommand \BibitemShut  [1]{\csname bibitem#1\endcsname}%
\let\auto@bib@innerbib\@empty
\bibitem [{\citenamefont {Bohm}\ and\ \citenamefont
  {Weinstein}(1948)}]{a1bohm}%
  \BibitemOpen
  \bibfield  {author} {\bibinfo {author} {\bibfnamefont {D.}~\bibnamefont
  {Bohm}}\ and\ \bibinfo {author} {\bibfnamefont {M.}~\bibnamefont
  {Weinstein}},\ }\bibfield  {title} {\bibinfo {title} {The self-oscillations
  of a charged particle},\ }\href@noop {} {\bibfield  {journal} {\bibinfo
  {journal} {Phys. Rev.}\ }\textbf {\bibinfo {volume} {74}},\ \bibinfo {pages}
  {1789} (\bibinfo {year} {1948})}\BibitemShut {NoStop}%
\bibitem [{\citenamefont {Goedecke}(1964)}]{a2goedecke}%
  \BibitemOpen
  \bibfield  {author} {\bibinfo {author} {\bibfnamefont {G.~H.}\ \bibnamefont
  {Goedecke}},\ }\bibfield  {title} {\bibinfo {title} {Classically
  radiationless motions and possible implications for quantum theory},\
  }\href@noop {} {\bibfield  {journal} {\bibinfo  {journal} {Phys. Rev.}\
  }\textbf {\bibinfo {volume} {135}},\ \bibinfo {pages} {B281} (\bibinfo {year}
  {1964})}\BibitemShut {NoStop}%
\bibitem [{\citenamefont {Zel’Dovich}(1958)}]{a3zel}%
  \BibitemOpen
  \bibfield  {author} {\bibinfo {author} {\bibfnamefont {I.~B.}\ \bibnamefont
  {Zel’Dovich}},\ }\bibfield  {title} {\bibinfo {title} {Electromagnetic
  interaction with parity violation},\ }\href@noop {} {\bibfield  {journal}
  {\bibinfo  {journal} {Sov. Phys. JETP}\ }\textbf {\bibinfo {volume} {6}},\
  \bibinfo {pages} {1184} (\bibinfo {year} {1958})}\BibitemShut {NoStop}%
\bibitem [{\citenamefont {Wood}\ \emph {et~al.}(1997)\citenamefont {Wood},
  \citenamefont {Bennett}, \citenamefont {Cho}, \citenamefont {Masterson},
  \citenamefont {Roberts}, \citenamefont {Tanner},\ and\ \citenamefont
  {Wieman}}]{a4wood}%
  \BibitemOpen
  \bibfield  {author} {\bibinfo {author} {\bibfnamefont {C.}~\bibnamefont
  {Wood}}, \bibinfo {author} {\bibfnamefont {S.}~\bibnamefont {Bennett}},
  \bibinfo {author} {\bibfnamefont {D.}~\bibnamefont {Cho}}, \bibinfo {author}
  {\bibfnamefont {B.}~\bibnamefont {Masterson}}, \bibinfo {author}
  {\bibfnamefont {J.}~\bibnamefont {Roberts}}, \bibinfo {author} {\bibfnamefont
  {C.}~\bibnamefont {Tanner}},\ and\ \bibinfo {author} {\bibfnamefont {C.~E.}\
  \bibnamefont {Wieman}},\ }\bibfield  {title} {\bibinfo {title} {Measurement
  of parity nonconservation and an anapole moment in cesium},\ }\href@noop {}
  {\bibfield  {journal} {\bibinfo  {journal} {Science}\ }\textbf {\bibinfo
  {volume} {275}},\ \bibinfo {pages} {1759} (\bibinfo {year}
  {1997})}\BibitemShut {NoStop}%
\bibitem [{\citenamefont {Yang}\ and\ \citenamefont
  {Bozhevolnyi}(2019)}]{b1yang}%
  \BibitemOpen
  \bibfield  {author} {\bibinfo {author} {\bibfnamefont {Y.}~\bibnamefont
  {Yang}}\ and\ \bibinfo {author} {\bibfnamefont {S.~I.}\ \bibnamefont
  {Bozhevolnyi}},\ }\bibfield  {title} {\bibinfo {title} {Nonradiating anapole
  states in nanophotonics: from fundamentals to applications},\ }\href@noop {}
  {\bibfield  {journal} {\bibinfo  {journal} {Nanotech.}\ }\textbf {\bibinfo
  {volume} {30}},\ \bibinfo {pages} {204001} (\bibinfo {year}
  {2019})}\BibitemShut {NoStop}%
\bibitem [{\citenamefont {Gurvitz}\ \emph {et~al.}(2019)\citenamefont
  {Gurvitz}, \citenamefont {Ladutenko}, \citenamefont {Dergachev},
  \citenamefont {Evlyukhin}, \citenamefont {Miroshnichenko},\ and\
  \citenamefont {Shalin}}]{a5gurvitz}%
  \BibitemOpen
  \bibfield  {author} {\bibinfo {author} {\bibfnamefont {E.~A.}\ \bibnamefont
  {Gurvitz}}, \bibinfo {author} {\bibfnamefont {K.~S.}\ \bibnamefont
  {Ladutenko}}, \bibinfo {author} {\bibfnamefont {P.~A.}\ \bibnamefont
  {Dergachev}}, \bibinfo {author} {\bibfnamefont {A.~B.}\ \bibnamefont
  {Evlyukhin}}, \bibinfo {author} {\bibfnamefont {A.~E.}\ \bibnamefont
  {Miroshnichenko}},\ and\ \bibinfo {author} {\bibfnamefont {A.~S.}\
  \bibnamefont {Shalin}},\ }\bibfield  {title} {\bibinfo {title} {The
  high-order toroidal moments and anapole states in all-dielectric photonics},\
  }\href@noop {} {\bibfield  {journal} {\bibinfo  {journal} {Laser \& Photon.
  Rev.}\ }\textbf {\bibinfo {volume} {13}},\ \bibinfo {pages} {1800266}
  (\bibinfo {year} {2019})}\BibitemShut {NoStop}%
\bibitem [{\citenamefont {Alaee}\ \emph {et~al.}(2018)\citenamefont {Alaee},
  \citenamefont {Rockstuhl},\ and\ \citenamefont
  {Fernandez-Corbaton}}]{a6alaee}%
  \BibitemOpen
  \bibfield  {author} {\bibinfo {author} {\bibfnamefont {R.}~\bibnamefont
  {Alaee}}, \bibinfo {author} {\bibfnamefont {C.}~\bibnamefont {Rockstuhl}},\
  and\ \bibinfo {author} {\bibfnamefont {I.}~\bibnamefont
  {Fernandez-Corbaton}},\ }\bibfield  {title} {\bibinfo {title} {An
  electromagnetic multipole expansion beyond the long-wavelength
  approximation},\ }\href@noop {} {\bibfield  {journal} {\bibinfo  {journal}
  {Opt. Commun.}\ }\textbf {\bibinfo {volume} {407}},\ \bibinfo {pages} {17}
  (\bibinfo {year} {2018})}\BibitemShut {NoStop}%
\bibitem [{\citenamefont {Kaelberer}\ \emph {et~al.}(2010)\citenamefont
  {Kaelberer}, \citenamefont {Fedotov}, \citenamefont {Papasimakis},
  \citenamefont {Tsai},\ and\ \citenamefont {Zheludev}}]{a7kael}%
  \BibitemOpen
  \bibfield  {author} {\bibinfo {author} {\bibfnamefont {T.}~\bibnamefont
  {Kaelberer}}, \bibinfo {author} {\bibfnamefont {V.}~\bibnamefont {Fedotov}},
  \bibinfo {author} {\bibfnamefont {N.}~\bibnamefont {Papasimakis}}, \bibinfo
  {author} {\bibfnamefont {D.}~\bibnamefont {Tsai}},\ and\ \bibinfo {author}
  {\bibfnamefont {N.}~\bibnamefont {Zheludev}},\ }\bibfield  {title} {\bibinfo
  {title} {Toroidal dipolar response in a metamaterial},\ }\href@noop {}
  {\bibfield  {journal} {\bibinfo  {journal} {Science}\ }\textbf {\bibinfo
  {volume} {330}},\ \bibinfo {pages} {1510} (\bibinfo {year}
  {2010})}\BibitemShut {NoStop}%
\bibitem [{\citenamefont {Fedotov}\ \emph {et~al.}(2013)\citenamefont
  {Fedotov}, \citenamefont {Rogacheva}, \citenamefont {Savinov}, \citenamefont
  {Tsai},\ and\ \citenamefont {Zheludev}}]{a8fedotov}%
  \BibitemOpen
  \bibfield  {author} {\bibinfo {author} {\bibfnamefont {V.~A.}\ \bibnamefont
  {Fedotov}}, \bibinfo {author} {\bibfnamefont {A.}~\bibnamefont {Rogacheva}},
  \bibinfo {author} {\bibfnamefont {V.}~\bibnamefont {Savinov}}, \bibinfo
  {author} {\bibfnamefont {D.~P.}\ \bibnamefont {Tsai}},\ and\ \bibinfo
  {author} {\bibfnamefont {N.~I.}\ \bibnamefont {Zheludev}},\ }\bibfield
  {title} {\bibinfo {title} {Resonant transparency and non-trivial
  non-radiating excitations in toroidal metamaterials},\ }\href@noop {}
  {\bibfield  {journal} {\bibinfo  {journal} {Sci. Reports}\ }\textbf {\bibinfo
  {volume} {3}},\ \bibinfo {pages} {2967} (\bibinfo {year} {2013})}\BibitemShut
  {NoStop}%
\bibitem [{\citenamefont {Miroshnichenko}\ \emph {et~al.}(2015)\citenamefont
  {Miroshnichenko}, \citenamefont {Evlyukhin}, \citenamefont {Yu},
  \citenamefont {Bakker}, \citenamefont {Chipouline}, \citenamefont
  {Kuznetsov}, \citenamefont {Luk’yanchuk}, \citenamefont {Chichkov},\ and\
  \citenamefont {Kivshar}}]{a9miro}%
  \BibitemOpen
  \bibfield  {author} {\bibinfo {author} {\bibfnamefont {A.~E.}\ \bibnamefont
  {Miroshnichenko}}, \bibinfo {author} {\bibfnamefont {A.~B.}\ \bibnamefont
  {Evlyukhin}}, \bibinfo {author} {\bibfnamefont {Y.~F.}\ \bibnamefont {Yu}},
  \bibinfo {author} {\bibfnamefont {R.~M.}\ \bibnamefont {Bakker}}, \bibinfo
  {author} {\bibfnamefont {A.}~\bibnamefont {Chipouline}}, \bibinfo {author}
  {\bibfnamefont {A.~I.}\ \bibnamefont {Kuznetsov}}, \bibinfo {author}
  {\bibfnamefont {B.}~\bibnamefont {Luk’yanchuk}}, \bibinfo {author}
  {\bibfnamefont {B.~N.}\ \bibnamefont {Chichkov}},\ and\ \bibinfo {author}
  {\bibfnamefont {Y.~S.}\ \bibnamefont {Kivshar}},\ }\bibfield  {title}
  {\bibinfo {title} {Nonradiating anapole modes in dielectric nanoparticles},\
  }\href@noop {} {\bibfield  {journal} {\bibinfo  {journal} {Nat. Commun.}\
  }\textbf {\bibinfo {volume} {6}},\ \bibinfo {pages} {8069} (\bibinfo {year}
  {2015})}\BibitemShut {NoStop}%
\bibitem [{\citenamefont {Kapitanova}\ \emph {et~al.}(2020)\citenamefont
  {Kapitanova}, \citenamefont {Zanganeh}, \citenamefont {Pavlov}, \citenamefont
  {Song}, \citenamefont {Belov}, \citenamefont {Evlyukhin},\ and\ \citenamefont
  {Miroshnichenko}}]{a11Kapitanova}%
  \BibitemOpen
  \bibfield  {author} {\bibinfo {author} {\bibfnamefont {P.}~\bibnamefont
  {Kapitanova}}, \bibinfo {author} {\bibfnamefont {E.}~\bibnamefont
  {Zanganeh}}, \bibinfo {author} {\bibfnamefont {N.}~\bibnamefont {Pavlov}},
  \bibinfo {author} {\bibfnamefont {M.}~\bibnamefont {Song}}, \bibinfo {author}
  {\bibfnamefont {P.}~\bibnamefont {Belov}}, \bibinfo {author} {\bibfnamefont
  {A.}~\bibnamefont {Evlyukhin}},\ and\ \bibinfo {author} {\bibfnamefont
  {A.}~\bibnamefont {Miroshnichenko}},\ }\bibfield  {title} {\bibinfo {title}
  {Seeing the unseen: experimental observation of magnetic anapole state inside
  a high-index dielectric particle},\ }\href@noop {} {\bibfield  {journal}
  {\bibinfo  {journal} {Annalen der Phys.}\ }\textbf {\bibinfo {volume}
  {532}},\ \bibinfo {pages} {2000293} (\bibinfo {year} {2020})}\BibitemShut
  {NoStop}%
\bibitem [{\citenamefont {Luk'yanchuk}\ \emph
  {et~al.}(2017{\natexlab{a}})\citenamefont {Luk'yanchuk}, \citenamefont
  {Paniagua-Dom{\'\i}nguez}, \citenamefont {Kuznetsov}, \citenamefont
  {Miroshnichenko},\ and\ \citenamefont {Kivshar}}]{a12luk}%
  \BibitemOpen
  \bibfield  {author} {\bibinfo {author} {\bibfnamefont {B.}~\bibnamefont
  {Luk'yanchuk}}, \bibinfo {author} {\bibfnamefont {R.}~\bibnamefont
  {Paniagua-Dom{\'\i}nguez}}, \bibinfo {author} {\bibfnamefont {A.~I.}\
  \bibnamefont {Kuznetsov}}, \bibinfo {author} {\bibfnamefont {A.~E.}\
  \bibnamefont {Miroshnichenko}},\ and\ \bibinfo {author} {\bibfnamefont
  {Y.~S.}\ \bibnamefont {Kivshar}},\ }\bibfield  {title} {\bibinfo {title}
  {Hybrid anapole modes of high-index dielectric nanoparticles},\ }\href@noop
  {} {\bibfield  {journal} {\bibinfo  {journal} {Phys. Rev. A}\ }\textbf
  {\bibinfo {volume} {95}},\ \bibinfo {pages} {063820} (\bibinfo {year}
  {2017}{\natexlab{a}})}\BibitemShut {NoStop}%
\bibitem [{\citenamefont {Luk'yanchuk}\ \emph
  {et~al.}(2017{\natexlab{b}})\citenamefont {Luk'yanchuk}, \citenamefont
  {Paniagua-Dom{\'\i}nguez}, \citenamefont {Kuznetsov}, \citenamefont
  {Miroshnichenko},\ and\ \citenamefont {Kivshar}}]{a13luk}%
  \BibitemOpen
  \bibfield  {author} {\bibinfo {author} {\bibfnamefont {B.}~\bibnamefont
  {Luk'yanchuk}}, \bibinfo {author} {\bibfnamefont {R.}~\bibnamefont
  {Paniagua-Dom{\'\i}nguez}}, \bibinfo {author} {\bibfnamefont {A.~I.}\
  \bibnamefont {Kuznetsov}}, \bibinfo {author} {\bibfnamefont {A.~E.}\
  \bibnamefont {Miroshnichenko}},\ and\ \bibinfo {author} {\bibfnamefont
  {Y.~S.}\ \bibnamefont {Kivshar}},\ }\bibfield  {title} {\bibinfo {title}
  {Suppression of scattering for small dielectric particles: anapole mode and
  invisibility},\ }\href@noop {} {\bibfield  {journal} {\bibinfo  {journal}
  {Phil. Tran. Royal Soc.: Math. Phys. Eng. Sci.}\ }\textbf {\bibinfo {volume}
  {375}},\ \bibinfo {pages} {20160069} (\bibinfo {year}
  {2017}{\natexlab{b}})}\BibitemShut {NoStop}%
\bibitem [{\citenamefont {Parker}\ \emph {et~al.}(2020)\citenamefont {Parker},
  \citenamefont {Sugimoto}, \citenamefont {Coe}, \citenamefont {Eggena},
  \citenamefont {Fujii}, \citenamefont {Scherer}, \citenamefont {Gray},\ and\
  \citenamefont {Manna}}]{a14parker}%
  \BibitemOpen
  \bibfield  {author} {\bibinfo {author} {\bibfnamefont {J.~A.}\ \bibnamefont
  {Parker}}, \bibinfo {author} {\bibfnamefont {H.}~\bibnamefont {Sugimoto}},
  \bibinfo {author} {\bibfnamefont {B.}~\bibnamefont {Coe}}, \bibinfo {author}
  {\bibfnamefont {D.}~\bibnamefont {Eggena}}, \bibinfo {author} {\bibfnamefont
  {M.}~\bibnamefont {Fujii}}, \bibinfo {author} {\bibfnamefont {N.~F.}\
  \bibnamefont {Scherer}}, \bibinfo {author} {\bibfnamefont {S.~K.}\
  \bibnamefont {Gray}},\ and\ \bibinfo {author} {\bibfnamefont
  {U.}~\bibnamefont {Manna}},\ }\bibfield  {title} {\bibinfo {title}
  {Excitation of nonradiating anapoles in dielectric nanospheres},\ }\href@noop
  {} {\bibfield  {journal} {\bibinfo  {journal} {Phys. Rev.Lett.}\ }\textbf
  {\bibinfo {volume} {124}},\ \bibinfo {pages} {097402} (\bibinfo {year}
  {2020})}\BibitemShut {NoStop}%
\bibitem [{\citenamefont {Zenin}\ \emph {et~al.}(2017)\citenamefont {Zenin},
  \citenamefont {Evlyukhin}, \citenamefont {Novikov}, \citenamefont {Yang},
  \citenamefont {Malureanu}, \citenamefont {Lavrinenko}, \citenamefont
  {Chichkov},\ and\ \citenamefont {Bozhevolnyi}}]{a15zenin}%
  \BibitemOpen
  \bibfield  {author} {\bibinfo {author} {\bibfnamefont {V.~A.}\ \bibnamefont
  {Zenin}}, \bibinfo {author} {\bibfnamefont {A.~B.}\ \bibnamefont
  {Evlyukhin}}, \bibinfo {author} {\bibfnamefont {S.~M.}\ \bibnamefont
  {Novikov}}, \bibinfo {author} {\bibfnamefont {Y.}~\bibnamefont {Yang}},
  \bibinfo {author} {\bibfnamefont {R.}~\bibnamefont {Malureanu}}, \bibinfo
  {author} {\bibfnamefont {A.~V.}\ \bibnamefont {Lavrinenko}}, \bibinfo
  {author} {\bibfnamefont {B.~N.}\ \bibnamefont {Chichkov}},\ and\ \bibinfo
  {author} {\bibfnamefont {S.~I.}\ \bibnamefont {Bozhevolnyi}},\ }\bibfield
  {title} {\bibinfo {title} {Direct amplitude-phase near-field observation of
  higher-order anapole states},\ }\href@noop {} {\bibfield  {journal} {\bibinfo
   {journal} {Nano Lett.}\ }\textbf {\bibinfo {volume} {17}},\ \bibinfo {pages}
  {7152} (\bibinfo {year} {2017})}\BibitemShut {NoStop}%
\bibitem [{\citenamefont {Zhang}\ \emph {et~al.}(2022)\citenamefont {Zhang},
  \citenamefont {Xue}, \citenamefont {Zhang}, \citenamefont {Li}, \citenamefont
  {Liu}, \citenamefont {Xie}, \citenamefont {Yao}, \citenamefont {Wang},
  \citenamefont {Ye},\ and\ \citenamefont {Zhu}}]{a16zhang}%
  \BibitemOpen
  \bibfield  {author} {\bibinfo {author} {\bibfnamefont {C.}~\bibnamefont
  {Zhang}}, \bibinfo {author} {\bibfnamefont {T.}~\bibnamefont {Xue}}, \bibinfo
  {author} {\bibfnamefont {J.}~\bibnamefont {Zhang}}, \bibinfo {author}
  {\bibfnamefont {Z.}~\bibnamefont {Li}}, \bibinfo {author} {\bibfnamefont
  {L.}~\bibnamefont {Liu}}, \bibinfo {author} {\bibfnamefont {J.}~\bibnamefont
  {Xie}}, \bibinfo {author} {\bibfnamefont {J.}~\bibnamefont {Yao}}, \bibinfo
  {author} {\bibfnamefont {G.}~\bibnamefont {Wang}}, \bibinfo {author}
  {\bibfnamefont {X.}~\bibnamefont {Ye}},\ and\ \bibinfo {author}
  {\bibfnamefont {W.}~\bibnamefont {Zhu}},\ }\bibfield  {title} {\bibinfo
  {title} {Terahertz meta-biosensor based on high-q electrical resonance
  enhanced by the interference of toroidal dipole},\ }\href@noop {} {\bibfield
  {journal} {\bibinfo  {journal} {Biosensors Bioelectron.}\ }\textbf {\bibinfo
  {volume} {214}},\ \bibinfo {pages} {114493} (\bibinfo {year}
  {2022})}\BibitemShut {NoStop}%
\bibitem [{\citenamefont {Zanganeh}\ \emph
  {et~al.}(2021{\natexlab{a}})\citenamefont {Zanganeh}, \citenamefont {Song},
  \citenamefont {Valero}, \citenamefont {Shalin}, \citenamefont {Nenasheva},
  \citenamefont {Miroshnichenko}, \citenamefont {Evlyukhin},\ and\
  \citenamefont {Kapitanova}}]{a17zanganeh}%
  \BibitemOpen
  \bibfield  {author} {\bibinfo {author} {\bibfnamefont {E.}~\bibnamefont
  {Zanganeh}}, \bibinfo {author} {\bibfnamefont {M.}~\bibnamefont {Song}},
  \bibinfo {author} {\bibfnamefont {A.~C.}\ \bibnamefont {Valero}}, \bibinfo
  {author} {\bibfnamefont {A.~S.}\ \bibnamefont {Shalin}}, \bibinfo {author}
  {\bibfnamefont {E.}~\bibnamefont {Nenasheva}}, \bibinfo {author}
  {\bibfnamefont {A.}~\bibnamefont {Miroshnichenko}}, \bibinfo {author}
  {\bibfnamefont {A.}~\bibnamefont {Evlyukhin}},\ and\ \bibinfo {author}
  {\bibfnamefont {P.}~\bibnamefont {Kapitanova}},\ }\bibfield  {title}
  {\bibinfo {title} {Nonradiating sources for efficient wireless power
  transfer},\ }\href@noop {} {\bibfield  {journal} {\bibinfo  {journal}
  {Nanophoton.}\ }\textbf {\bibinfo {volume} {10}},\ \bibinfo {pages} {4399}
  (\bibinfo {year} {2021}{\natexlab{a}})}\BibitemShut {NoStop}%
\bibitem [{\citenamefont {Savinov}\ \emph {et~al.}(2019)\citenamefont
  {Savinov}, \citenamefont {Papasimakis}, \citenamefont {Tsai},\ and\
  \citenamefont {Zheludev}}]{a18savinov}%
  \BibitemOpen
  \bibfield  {author} {\bibinfo {author} {\bibfnamefont {V.}~\bibnamefont
  {Savinov}}, \bibinfo {author} {\bibfnamefont {N.}~\bibnamefont
  {Papasimakis}}, \bibinfo {author} {\bibfnamefont {D.}~\bibnamefont {Tsai}},\
  and\ \bibinfo {author} {\bibfnamefont {N.}~\bibnamefont {Zheludev}},\
  }\bibfield  {title} {\bibinfo {title} {Optical anapoles},\ }\href@noop {}
  {\bibfield  {journal} {\bibinfo  {journal} {Commun. Phys.}\ }\textbf
  {\bibinfo {volume} {2}},\ \bibinfo {pages} {69} (\bibinfo {year}
  {2019})}\BibitemShut {NoStop}%
\bibitem [{\citenamefont {Nemkov}\ \emph {et~al.}(2017)\citenamefont {Nemkov},
  \citenamefont {Stenishchev},\ and\ \citenamefont {Basharin}}]{a19nemkov}%
  \BibitemOpen
  \bibfield  {author} {\bibinfo {author} {\bibfnamefont {N.~A.}\ \bibnamefont
  {Nemkov}}, \bibinfo {author} {\bibfnamefont {I.~V.}\ \bibnamefont
  {Stenishchev}},\ and\ \bibinfo {author} {\bibfnamefont {A.~A.}\ \bibnamefont
  {Basharin}},\ }\bibfield  {title} {\bibinfo {title} {Nontrivial nonradiating
  all-dielectric anapole},\ }\href@noop {} {\bibfield  {journal} {\bibinfo
  {journal} {Sci. Reports}\ }\textbf {\bibinfo {volume} {7}},\ \bibinfo {pages}
  {1064} (\bibinfo {year} {2017})}\BibitemShut {NoStop}%
\bibitem [{\citenamefont {Zanganeh}\ \emph
  {et~al.}(2021{\natexlab{b}})\citenamefont {Zanganeh}, \citenamefont
  {Evlyukhin}, \citenamefont {Miroshnichenko}, \citenamefont {Song},
  \citenamefont {Nenasheva},\ and\ \citenamefont {Kapitanova}}]{a20zanganeh}%
  \BibitemOpen
  \bibfield  {author} {\bibinfo {author} {\bibfnamefont {E.}~\bibnamefont
  {Zanganeh}}, \bibinfo {author} {\bibfnamefont {A.}~\bibnamefont {Evlyukhin}},
  \bibinfo {author} {\bibfnamefont {A.}~\bibnamefont {Miroshnichenko}},
  \bibinfo {author} {\bibfnamefont {M.}~\bibnamefont {Song}}, \bibinfo {author}
  {\bibfnamefont {E.}~\bibnamefont {Nenasheva}},\ and\ \bibinfo {author}
  {\bibfnamefont {P.}~\bibnamefont {Kapitanova}},\ }\bibfield  {title}
  {\bibinfo {title} {Anapole meta-atoms: nonradiating electric and magnetic
  sources},\ }\href@noop {} {\bibfield  {journal} {\bibinfo  {journal} {Phys.
  Rev. Lett.}\ }\textbf {\bibinfo {volume} {127}},\ \bibinfo {pages} {096804}
  (\bibinfo {year} {2021}{\natexlab{b}})}\BibitemShut {NoStop}%
\bibitem [{\citenamefont {Zanganeh}\ \emph
  {et~al.}(2021{\natexlab{c}})\citenamefont {Zanganeh}, \citenamefont {Song},
  \citenamefont {Valero}, \citenamefont {Shalin}, \citenamefont {Nenasheva},
  \citenamefont {Miroshnichenko}, \citenamefont {Evlyukhin},\ and\
  \citenamefont {Kapitanova}}]{a21zanganeh}%
  \BibitemOpen
  \bibfield  {author} {\bibinfo {author} {\bibfnamefont {E.}~\bibnamefont
  {Zanganeh}}, \bibinfo {author} {\bibfnamefont {M.}~\bibnamefont {Song}},
  \bibinfo {author} {\bibfnamefont {A.~C.}\ \bibnamefont {Valero}}, \bibinfo
  {author} {\bibfnamefont {A.~S.}\ \bibnamefont {Shalin}}, \bibinfo {author}
  {\bibfnamefont {E.}~\bibnamefont {Nenasheva}}, \bibinfo {author}
  {\bibfnamefont {A.}~\bibnamefont {Miroshnichenko}}, \bibinfo {author}
  {\bibfnamefont {A.}~\bibnamefont {Evlyukhin}},\ and\ \bibinfo {author}
  {\bibfnamefont {P.}~\bibnamefont {Kapitanova}},\ }\bibfield  {title}
  {\bibinfo {title} {Nonradiating sources for efficient wireless power
  transfer},\ }\href@noop {} {\bibfield  {journal} {\bibinfo  {journal}
  {Nanophoton.}\ }\textbf {\bibinfo {volume} {10}},\ \bibinfo {pages} {4399}
  (\bibinfo {year} {2021}{\natexlab{c}})}\BibitemShut {NoStop}%
\bibitem [{\citenamefont {Zurita-S{\'a}nchez}(2019)}]{a22zurita}%
  \BibitemOpen
  \bibfield  {author} {\bibinfo {author} {\bibfnamefont {J.~R.}\ \bibnamefont
  {Zurita-S{\'a}nchez}},\ }\bibfield  {title} {\bibinfo {title} {Anapole
  arising from a mie scatterer with dipole excitation},\ }\href@noop {}
  {\bibfield  {journal} {\bibinfo  {journal} {Phys. Rev. Research}\ }\textbf
  {\bibinfo {volume} {1}},\ \bibinfo {pages} {033064} (\bibinfo {year}
  {2019})}\BibitemShut {NoStop}%
\bibitem [{\citenamefont {Evlyukhin}\ \emph {et~al.}(2016)\citenamefont
  {Evlyukhin}, \citenamefont {Fischer}, \citenamefont {Reinhardt},\ and\
  \citenamefont {Chichkov}}]{a23evlyukhin}%
  \BibitemOpen
  \bibfield  {author} {\bibinfo {author} {\bibfnamefont {A.~B.}\ \bibnamefont
  {Evlyukhin}}, \bibinfo {author} {\bibfnamefont {T.}~\bibnamefont {Fischer}},
  \bibinfo {author} {\bibfnamefont {C.}~\bibnamefont {Reinhardt}},\ and\
  \bibinfo {author} {\bibfnamefont {B.~N.}\ \bibnamefont {Chichkov}},\
  }\bibfield  {title} {\bibinfo {title} {Optical theorem and multipole
  scattering of light by arbitrarily shaped nanoparticles},\ }\href@noop {}
  {\bibfield  {journal} {\bibinfo  {journal} {Phys. Rev. B}\ }\textbf {\bibinfo
  {volume} {94}},\ \bibinfo {pages} {205434} (\bibinfo {year}
  {2016})}\BibitemShut {NoStop}%
\bibitem [{a36()}]{a36}%
  \BibitemOpen
  \href@noop {} {}\bibinfo {howpublished} {See Supplemental Material
  \url{url}},\ \bibinfo {note} {for calculations and more details on design,
  which includes Refs. [38-39]}\BibitemShut {NoStop}%
\bibitem [{\citenamefont {Monticone}\ \emph {et~al.}(2019)\citenamefont
  {Monticone}, \citenamefont {Sounas}, \citenamefont {Krasnok},\ and\
  \citenamefont {Al{\`u}}}]{a24mon}%
  \BibitemOpen
  \bibfield  {author} {\bibinfo {author} {\bibfnamefont {F.}~\bibnamefont
  {Monticone}}, \bibinfo {author} {\bibfnamefont {D.}~\bibnamefont {Sounas}},
  \bibinfo {author} {\bibfnamefont {A.}~\bibnamefont {Krasnok}},\ and\ \bibinfo
  {author} {\bibfnamefont {A.}~\bibnamefont {Al{\`u}}},\ }\bibfield  {title}
  {\bibinfo {title} {Can a nonradiating mode be externally excited?
  nonscattering states versus embedded eigenstates},\ }\href@noop {} {\bibfield
   {journal} {\bibinfo  {journal} {ACS Photon.}\ }\textbf {\bibinfo {volume}
  {6}},\ \bibinfo {pages} {3108} (\bibinfo {year} {2019})}\BibitemShut
  {NoStop}%
\bibitem [{\citenamefont {Hsu}\ \emph {et~al.}(2016)\citenamefont {Hsu},
  \citenamefont {Zhen}, \citenamefont {Stone}, \citenamefont {Joannopoulos},\
  and\ \citenamefont {Solja{\v{c}}i{\'c}}}]{a25hsu}%
  \BibitemOpen
  \bibfield  {author} {\bibinfo {author} {\bibfnamefont {C.~W.}\ \bibnamefont
  {Hsu}}, \bibinfo {author} {\bibfnamefont {B.}~\bibnamefont {Zhen}}, \bibinfo
  {author} {\bibfnamefont {A.~D.}\ \bibnamefont {Stone}}, \bibinfo {author}
  {\bibfnamefont {J.~D.}\ \bibnamefont {Joannopoulos}},\ and\ \bibinfo {author}
  {\bibfnamefont {M.}~\bibnamefont {Solja{\v{c}}i{\'c}}},\ }\bibfield  {title}
  {\bibinfo {title} {Bound states in the continuum},\ }\href@noop {} {\bibfield
   {journal} {\bibinfo  {journal} {Nat. Rev. Mater.}\ }\textbf {\bibinfo
  {volume} {1}},\ \bibinfo {pages} {1} (\bibinfo {year} {2016})}\BibitemShut
  {NoStop}%
\bibitem [{\citenamefont {Monticone}\ and\ \citenamefont {Alu}(2014)}]{a26mon}%
  \BibitemOpen
  \bibfield  {author} {\bibinfo {author} {\bibfnamefont {F.}~\bibnamefont
  {Monticone}}\ and\ \bibinfo {author} {\bibfnamefont {A.}~\bibnamefont
  {Alu}},\ }\bibfield  {title} {\bibinfo {title} {Embedded photonic eigenvalues
  in 3d nanostructures},\ }\href@noop {} {\bibfield  {journal} {\bibinfo
  {journal} {Phys. Rev. Lett.}\ }\textbf {\bibinfo {volume} {112}},\ \bibinfo
  {pages} {213903} (\bibinfo {year} {2014})}\BibitemShut {NoStop}%
\bibitem [{\citenamefont {Hsu}\ \emph {et~al.}(2013)\citenamefont {Hsu},
  \citenamefont {Zhen}, \citenamefont {Lee}, \citenamefont {Chua},
  \citenamefont {Johnson}, \citenamefont {Joannopoulos},\ and\ \citenamefont
  {Solja{\v{c}}i{\'c}}}]{a27hsu}%
  \BibitemOpen
  \bibfield  {author} {\bibinfo {author} {\bibfnamefont {C.~W.}\ \bibnamefont
  {Hsu}}, \bibinfo {author} {\bibfnamefont {B.}~\bibnamefont {Zhen}}, \bibinfo
  {author} {\bibfnamefont {J.}~\bibnamefont {Lee}}, \bibinfo {author}
  {\bibfnamefont {S.-L.}\ \bibnamefont {Chua}}, \bibinfo {author}
  {\bibfnamefont {S.~G.}\ \bibnamefont {Johnson}}, \bibinfo {author}
  {\bibfnamefont {J.~D.}\ \bibnamefont {Joannopoulos}},\ and\ \bibinfo {author}
  {\bibfnamefont {M.}~\bibnamefont {Solja{\v{c}}i{\'c}}},\ }\bibfield  {title}
  {\bibinfo {title} {Observation of trapped light within the radiation
  continuum},\ }\href@noop {} {\bibfield  {journal} {\bibinfo  {journal}
  {Nat.}\ }\textbf {\bibinfo {volume} {499}},\ \bibinfo {pages} {188} (\bibinfo
  {year} {2013})}\BibitemShut {NoStop}%
\bibitem [{\citenamefont {Silveirinha}(2014)}]{a28silv}%
  \BibitemOpen
  \bibfield  {author} {\bibinfo {author} {\bibfnamefont {M.~G.}\ \bibnamefont
  {Silveirinha}},\ }\bibfield  {title} {\bibinfo {title} {Trapping light in
  open plasmonic nanostructures},\ }\href@noop {} {\bibfield  {journal}
  {\bibinfo  {journal} {Phys. Rev. A}\ }\textbf {\bibinfo {volume} {89}},\
  \bibinfo {pages} {023813} (\bibinfo {year} {2014})}\BibitemShut {NoStop}%
\bibitem [{\citenamefont {Monticone}\ \emph {et~al.}(2018)\citenamefont
  {Monticone}, \citenamefont {Doeleman}, \citenamefont {Den~Hollander},
  \citenamefont {Koenderink},\ and\ \citenamefont {Al{\`u}}}]{a29mon}%
  \BibitemOpen
  \bibfield  {author} {\bibinfo {author} {\bibfnamefont {F.}~\bibnamefont
  {Monticone}}, \bibinfo {author} {\bibfnamefont {H.~M.}\ \bibnamefont
  {Doeleman}}, \bibinfo {author} {\bibfnamefont {W.}~\bibnamefont
  {Den~Hollander}}, \bibinfo {author} {\bibfnamefont {A.~F.}\ \bibnamefont
  {Koenderink}},\ and\ \bibinfo {author} {\bibfnamefont {A.}~\bibnamefont
  {Al{\`u}}},\ }\bibfield  {title} {\bibinfo {title} {Trapping light in plain
  sight: Embedded photonic eigenstates in zero-index metamaterials},\
  }\href@noop {} {\bibfield  {journal} {\bibinfo  {journal} {Laser \& Photon.
  Rev.}\ }\textbf {\bibinfo {volume} {12}},\ \bibinfo {pages} {1700220}
  (\bibinfo {year} {2018})}\BibitemShut {NoStop}%
\bibitem [{\citenamefont {Doeleman}\ \emph {et~al.}(2018)\citenamefont
  {Doeleman}, \citenamefont {Monticone}, \citenamefont {den Hollander},
  \citenamefont {Al{\`u}},\ and\ \citenamefont {Koenderink}}]{a30doel}%
  \BibitemOpen
  \bibfield  {author} {\bibinfo {author} {\bibfnamefont {H.~M.}\ \bibnamefont
  {Doeleman}}, \bibinfo {author} {\bibfnamefont {F.}~\bibnamefont {Monticone}},
  \bibinfo {author} {\bibfnamefont {W.}~\bibnamefont {den Hollander}}, \bibinfo
  {author} {\bibfnamefont {A.}~\bibnamefont {Al{\`u}}},\ and\ \bibinfo {author}
  {\bibfnamefont {A.~F.}\ \bibnamefont {Koenderink}},\ }\bibfield  {title}
  {\bibinfo {title} {Experimental observation of a polarization vortex at an
  optical bound state in the continuum},\ }\href@noop {} {\bibfield  {journal}
  {\bibinfo  {journal} {Nat. Photon.}\ }\textbf {\bibinfo {volume} {12}},\
  \bibinfo {pages} {397} (\bibinfo {year} {2018})}\BibitemShut {NoStop}%
\bibitem [{\citenamefont {Al{\`u}}\ and\ \citenamefont
  {Engheta}(2005)}]{a31alu}%
  \BibitemOpen
  \bibfield  {author} {\bibinfo {author} {\bibfnamefont {A.}~\bibnamefont
  {Al{\`u}}}\ and\ \bibinfo {author} {\bibfnamefont {N.}~\bibnamefont
  {Engheta}},\ }\bibfield  {title} {\bibinfo {title} {Achieving transparency
  with plasmonic and metamaterial coatings},\ }\href@noop {} {\bibfield
  {journal} {\bibinfo  {journal} {Phys. Rev. E}\ }\textbf {\bibinfo {volume}
  {72}},\ \bibinfo {pages} {016623} (\bibinfo {year} {2005})}\BibitemShut
  {NoStop}%
\bibitem [{\citenamefont {Al{\`u}}\ and\ \citenamefont
  {Engheta}(2008)}]{a32alu}%
  \BibitemOpen
  \bibfield  {author} {\bibinfo {author} {\bibfnamefont {A.}~\bibnamefont
  {Al{\`u}}}\ and\ \bibinfo {author} {\bibfnamefont {N.}~\bibnamefont
  {Engheta}},\ }\bibfield  {title} {\bibinfo {title} {Multifrequency optical
  invisibility cloak with layered plasmonic shells},\ }\href@noop {} {\bibfield
   {journal} {\bibinfo  {journal} {Phys. Rev. Lett.}\ }\textbf {\bibinfo
  {volume} {100}},\ \bibinfo {pages} {113901} (\bibinfo {year}
  {2008})}\BibitemShut {NoStop}%
\bibitem [{\citenamefont {Edwards}\ \emph {et~al.}(2009)\citenamefont
  {Edwards}, \citenamefont {Al{\`u}}, \citenamefont {Silveirinha},\ and\
  \citenamefont {Engheta}}]{a33edwards}%
  \BibitemOpen
  \bibfield  {author} {\bibinfo {author} {\bibfnamefont {B.}~\bibnamefont
  {Edwards}}, \bibinfo {author} {\bibfnamefont {A.}~\bibnamefont {Al{\`u}}},
  \bibinfo {author} {\bibfnamefont {M.~G.}\ \bibnamefont {Silveirinha}},\ and\
  \bibinfo {author} {\bibfnamefont {N.}~\bibnamefont {Engheta}},\ }\bibfield
  {title} {\bibinfo {title} {Experimental verification of plasmonic cloaking at
  microwave frequencies with metamaterials},\ }\href@noop {} {\bibfield
  {journal} {\bibinfo  {journal} {Phys. Rev. Lett.}\ }\textbf {\bibinfo
  {volume} {103}},\ \bibinfo {pages} {153901} (\bibinfo {year}
  {2009})}\BibitemShut {NoStop}%
\bibitem [{\citenamefont {Al{\`u}}\ and\ \citenamefont
  {Engheta}(2010)}]{a34alu}%
  \BibitemOpen
  \bibfield  {author} {\bibinfo {author} {\bibfnamefont {A.}~\bibnamefont
  {Al{\`u}}}\ and\ \bibinfo {author} {\bibfnamefont {N.}~\bibnamefont
  {Engheta}},\ }\bibfield  {title} {\bibinfo {title} {Cloaked near-field
  scanning optical microscope tip for noninvasive near-field imaging},\
  }\href@noop {} {\bibfield  {journal} {\bibinfo  {journal} {Phys. Rev. Lett.}\
  }\textbf {\bibinfo {volume} {105}},\ \bibinfo {pages} {263906} (\bibinfo
  {year} {2010})}\BibitemShut {NoStop}%
\bibitem [{\citenamefont {Bilotti}\ \emph {et~al.}(2011)\citenamefont
  {Bilotti}, \citenamefont {Tricarico}, \citenamefont {Pierini},\ and\
  \citenamefont {Vegni}}]{a35bilotti}%
  \BibitemOpen
  \bibfield  {author} {\bibinfo {author} {\bibfnamefont {F.}~\bibnamefont
  {Bilotti}}, \bibinfo {author} {\bibfnamefont {S.}~\bibnamefont {Tricarico}},
  \bibinfo {author} {\bibfnamefont {F.}~\bibnamefont {Pierini}},\ and\ \bibinfo
  {author} {\bibfnamefont {L.}~\bibnamefont {Vegni}},\ }\bibfield  {title}
  {\bibinfo {title} {Cloaking apertureless near-field scanning optical
  microscopy tips},\ }\href@noop {} {\bibfield  {journal} {\bibinfo  {journal}
  {Opt. Lett.}\ }\textbf {\bibinfo {volume} {36}},\ \bibinfo {pages} {211}
  (\bibinfo {year} {2011})}\BibitemShut {NoStop}%
\bibitem [{\citenamefont {Petosa}(2007)}]{a37}%
  \BibitemOpen
  \bibfield  {author} {\bibinfo {author} {\bibfnamefont {A.}~\bibnamefont
  {Petosa}},\ }\href@noop {} {\emph {\bibinfo {title} {Dielectric resonator
  antenna handbook}}}\ (\bibinfo  {publisher} {Artech},\ \bibinfo {year}
  {2007})\BibitemShut {NoStop}%
\bibitem [{\citenamefont {Mongia}\ and\ \citenamefont {Bhartia}(1994)}]{a38}%
  \BibitemOpen
  \bibfield  {author} {\bibinfo {author} {\bibfnamefont {R.~K.}\ \bibnamefont
  {Mongia}}\ and\ \bibinfo {author} {\bibfnamefont {P.}~\bibnamefont
  {Bhartia}},\ }\bibfield  {title} {\bibinfo {title} {Dielectric resonator
  antennas—a review and general design relations for resonant frequency and
  bandwidth},\ }\href@noop {} {\bibfield  {journal} {\bibinfo  {journal}
  {International J. Microwav. Millimeter-Wav. Comp.-Aided Eng.}\ }\textbf
  {\bibinfo {volume} {4}},\ \bibinfo {pages} {230} (\bibinfo {year}
  {1994})}\BibitemShut {NoStop}%
\end{thebibliography}%

\end{document}